\documentclass[preprint]{jpsj3}
\usepackage{color}
\usepackage{bm}
\usepackage{mathrsfs}

\title{Ground States of Spin-1/2 Heisenberg Antiferromagnets with Frustration on a Diamond-Like Decorated Square Lattice}
\author{Yuhei  Hirose, Akihide Oguchi, and Yoshiyuki Fukumoto}
\inst{Department of Physics, Faculty of Science and Technology,\\ Tokyo University of Science, Noda, Chiba 278-8510}
\recdate{\today}
\abst{We study the ground-state phase diagram of a Heisenberg model with spin $S=\frac{1}{2}$ on a diamond-like decorated square lattice. 
A diamond unit has two types of antiferromagnetic exchange interactions, and the ratio $\lambda$ between the length of the diagonal bond and that of the  other four edges determines the strength of frustration.
It has been pointed out [J. Phys. Soc. Jpn {\bf85}, 033705 (2016)] that the so-called tetramer-dimer states, which are expected to be stabilized in an intermediate region of $\lambda_{\rm c}<\lambda<2$, 
are identical to the square-lattice dimer covering states, which ignited renewed interest in high-dimensional diamond-like decorated lattices. 
In order to determine the phase boundary $\lambda_{\rm c}$, we employ the modified spin wave method to estimate the energy of the ferrimagnetic state and obtain $\lambda_{\rm c}=0.974$.
Our obtained magnetizations for spin-$\frac{1}{2}$ sites and for spin-1 sites are $m=0.398$ and $\tilde{m}=0.949$, and spin reductions are 20 \% and 5\%, respectively. This indicates that spin fluctuation is much smaller than that of the $S=\frac{1}{2}$ square-lattice antiferromagnet: thus, we can consider that our obtained ground-state energy is highly accurate. 
Further, our numerical diagonalization study suggests that other cluster states do not appear in the ground-state phase diagram.
}

\begin{document}
\maketitle
\section{Introduction}\label{sec:1}
The exploration of frustration in quantum spin models has been one of the most interesting issues in condensed matter physics\cite{ref0}. 
The systems consisting of diamond units with frustration have also attracted wide attention both experimentally and theoretically. \cite{ref1,ref2,ref3-0,ref3,ref4,ref5,ref6,ref7-1,ref7-2}

\begin{figure}[h] 
\begin{center}
\includegraphics[width=.75\linewidth]{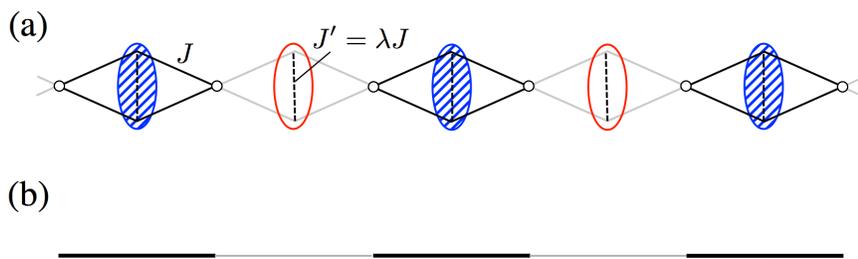}
\end{center}
\caption{(a) The tetramer-dimer (TD) state for $0.909<\lambda<2$ on a diamond chain. 
(b) Dimer model on a linear chain. The shaded blue and unshaded red ovals represent the triplet and singlet pairs, respectively. 
The small open circles in (a), which denote edge spins, always belong to one tetramer. 
The thick lines in (b) correspond to the tetramer on the diamond chain.
}
\label{fig:0}
\end{figure}

The diamond chain is one of the typical systems with diamond units, and it was proposed by  Takano {\it et al}.\cite{ref1}
It is a one-dimensional lattice system, and a diamond unit has two types of antiferromagnetic interactions. 
As shown in Fig.~{\ref{fig:0}} (a), solid and dashed lines, respectively, represent exchange parameters $J$ and $J'$, 
and the ratio $\lambda=J'/J$ determines the ground-state properties. 
It has been known that, in the case of spin $S=\frac{1}{2}$, three types of ground-state phases exist: the dimer-monomer (DM) state for $2<\lambda$, the tetramer-dimer (TD) state for $0.909<\lambda<2$, and the ferrimagnetic state for $\lambda<0.909$.\cite{ref1}
In the TD state, as shown in Fig.~{\ref{fig:0}} (a), diamond units with triplet pairs (shaded blue ovals) and with singlet pairs (unshaded red ovals) are arranged alternately.
This arrangement results from the fact that for $\lambda < 2$, the energy decreases as the number of triplet pairs increases, but nearest-neighbor repulsion exists between two diamond units with triplet pairs, which can be explained by the variational principle and the Lieb-Mattice theorem. \cite{ref1,ref2}
In the TD state, the edge spins, which are represented by the small open circles in Fig.~{\ref{fig:0}} (a), always belong to a tetramer.
Furthermore, as we will show later, the singlet pair on the dotted lines makes the four interactions $J$ in the diamond unit vanish effectively.
We note that the above-mentioned property of a tetramer is that of a dimer in the dimer covering model.
If we regard a tetramer as a ``dimer'', we can identify the doubly degenerate TD configuration with the doubly degenerate dimer covering states on a linear chain, as shown in Fig.~{\ref{fig:0}}~(b).
The singlet dimer, which takes the central role in ordinary RVB physics, has orientation and does not have orthogonality. 
On the other hand, our dimer coverings have orthogonality and do not have orientation.  

\begin{figure}[h] 
\begin{center}
\includegraphics[width=.40\linewidth]{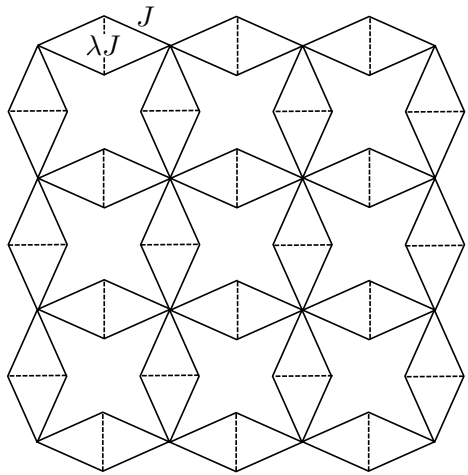}
\end{center}
\caption{Structure of a diamond-like decorated square lattice.  Solid and dotted lines represent the antiferromagnetic interactions, $J$ and $\lambda J$, respectively.
 }
\label{fig:1}
\end{figure}

\begin{figure}[!htb] 
\begin{center}
\includegraphics[width=.75\linewidth]{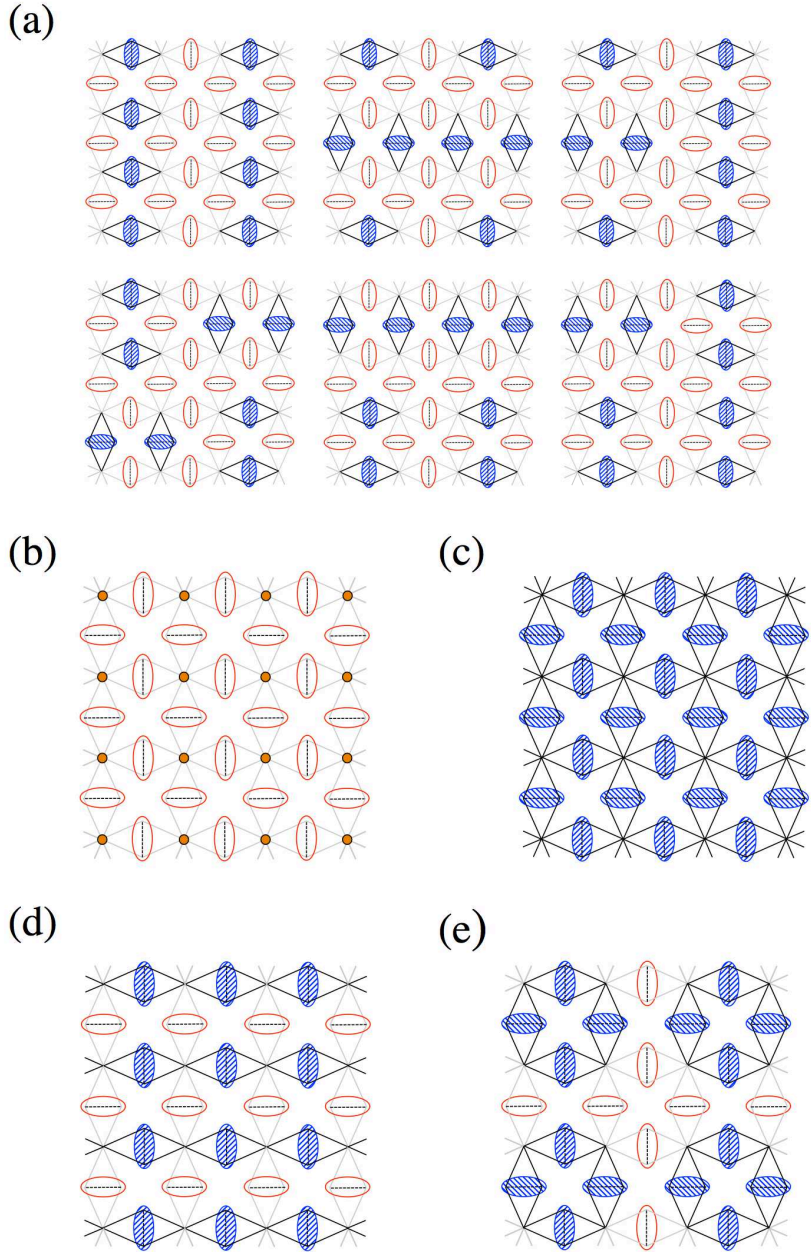}
\end{center}
\caption{(Color online) Arrangement of ground-state phases, (a) MDTD state for $0.974<\lambda<2$, (b) DM state for $2<\lambda$, and (c) ferrimagnetic state for $\lambda<0.974$. Arrangement of (d) chain ferrimagnetic and (e) square ferrimagnetic state phases. The orange circles of Fig. (b) represent the free spins. Chain and square ferrimagnetic states are not the ground states.
 }
\label{fig:2}
\end{figure}

The fact that a TD state is identical to a dimer covering state leads us toward the consideration of higher-dimensional systems.
When we extend the diamond chain to two dimensions or replace the solid lines of the square lattice with diamond units as shown in Fig.~{\ref{fig:1}}, which shows a lattice called diamond-like decorated square lattice\cite{ref7-1,ref7-2}, the TD phase shows very interesting phenomena, such as the emergence of the square-lattice dimer model. Because there are many kinds of TD states, or dimer-covering states, as shown in Fig.~{\ref{fig:2}} (a), the present system has a nontrivial macroscopic degeneracy. 
This idea has already been pointed out by Morita and Shibata,\cite{ref8} and the present authors have derived a quantum dimer model (QDM) as the second-order effective Hamiltonian for the model with further-neighbor couplings.\cite{ref9}

We call the TD state on the diamond-like decorated square lattice the ``macroscopically degenerated tetramer dimer (MDTD) state''. 
The wave function of a MDTD state is a direct product of the tetramer-singlet and dimer-singlet states, and it is a non-magnetic ground state. 
Although the MDTD states are expected to be ground-state manifolds in an intermediate region of $\lambda_{\rm c}<\lambda<2$, 
the phase boundary $\lambda_{\rm c}$ has not been determined yet.
The estimation of $\lambda_{\rm c}$ is one of the purposes of this paper.

We obtain $\lambda_{\rm c}$, which is assumed to be the intersection of MDTD and ferrimagnetic-state energies. 
By means of the modified spin wave method, we calculate the ferrimagnetic ground-state energy and obtain $\lambda_{\rm c}=0.974$.
Our obtained ground-state phases in the diamond-like decorated square lattice are as follows. 
For $2<\lambda$, we obtain the DM state, which has dimers (singlet pairs) and monomers (free spins) as shown in Fig.~{\ref{fig:2}} (b). 
For $\lambda<\lambda_{\rm c}$, we obtain the ferrimagnetic state as shown in Fig.~{\ref{fig:2}} (c). For $\lambda_{\rm c}<\lambda<2$, the ground state is the MDTD state.

Here, one may question whether there are other favored candidates for the ground state.
Thus, we investigate other arrangements as shown in Figs.~{\ref{fig:2}} (d) and (e). 
We refer to the arrangements of Figs.~{\ref{fig:2}} (d) and (e) as ``chain ferrimagnetic  state'', where diamonds with triplet pairs spread one-dimensionally, and ``square ferrimagnetic state'', where they spread two-dimensionally.
We find that both states are not the ground states, which suggests that there appear only three types of ground states in a diamond-like decorated square lattice. 
Although new types of ground states will be found in the future,  we pursue the subject under the assumption that number of ground states is three.

The remainder of this paper is organized as follows.
The Heisenberg model and the partition function of the diamond-like decorated square lattice are defined in Sect. 2. 
In Sect. 3, we calculate the ground-state energies of each state and determine the phase boundaries. 
In Sect. 4, we show the dependence of the energy on the parameter $\lambda$ and the ground-state phase diagram. 
In Sect. 7, we summarize the results obtained in this study.

\section{Hamiltonian and Partition Function}\label{sec:2}

\begin{figure}[h]
\begin{center}
\includegraphics[width=.40\linewidth]{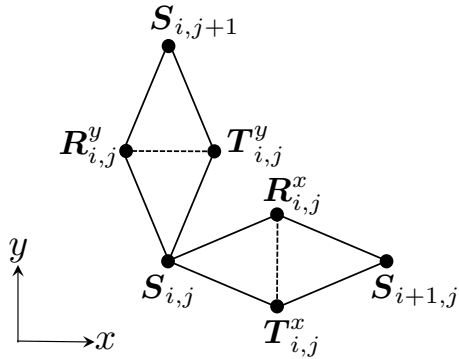}
\end{center}
\caption{Unit cell of a diamond-like decorated square lattice. }
\label{fig:4}
\end{figure}

We consider the unit cell shown in Fig.~{\ref{fig:4}}. 
Its Hamiltonian can be written as
\begin{align}
\mathcal{H}= \sum_{i,j} (\mathcal{H}_{i,j}+\lambda\mathcal{H}_{i,j}^{\lambda}),
\label{eq:1}
\end{align}
where $(i,j)$ denotes a lattice point, the summation is taken over all the lattice points, and 
\begin{align}
\mathcal{H}_{i,j}=J\left\{\left(\mbox{\boldmath $S$}_{i,j}+\mbox{\boldmath $S$}_{i+1,j}\right)\cdot\left(\mbox{\boldmath $R$}_{i,j}^x+\mbox{\boldmath $T$}_{i,j}^x\right)+\left(\mbox{\boldmath $S$}_{i,j}+\mbox{\boldmath $S$}_{i,j+1}\right)\cdot\left(\mbox{\boldmath $R$}_{i,j}^y+\mbox{\boldmath $T$}_{i,j}^y\right)\right\},
\label{eq:2}
\end{align}
\begin{align}
\mathcal{H}_{i,j}^{\lambda}=J\sum_{\tau=x,y}\left(\bm{R}_{i,j}^{\tau}\cdot\bm{T}_{i,j}^{\tau}+\frac{3}{4}\right),
\label{eq:3}
\end{align}
with spin $\frac{1}{2}$ operators $\mbox{\boldmath $S$}_{i,j}$, $\bm{R}_{i,j}^{\tau}$, and $\bm{T}_{i,j}^{\tau}$ $(\tau=x,y)$. 
In Eq.~(\ref{eq:3}), note that the energy of a dashed bond is measured from that of the singlet dimer.
Because $\left[\mathcal{H}_{i,j},\mathcal{H}_{i,j}^{\lambda}\right]=0$, $|\bm{R}_{i,j}^{\tau}+\bm{T}_{i,j}^{\tau}|$ are conserved quantities. 

Using $\left[\mathcal{H}_{i,j},\mathcal{H}_{i,j}^{\lambda}\right]=0$, we can express the partition function as 
\begin{align}
Z=&\mathsf{Tr}\left(e^{-\beta\sum_{i,j}\mathcal{H}_{i,j}}\cdot e^{-\beta\lambda\sum_{i,j}\mathcal{H}_{i,j}^{\lambda}}\right) \notag \\
       =&\mathsf{Tr}\Biggr[\prod_{i,j}e^{-K {\left(\bm{S}_{i,j}+\bm{S}_{i+1,j}\right)\cdot \bm{X}_{i,j}}+{\left(\bm{S}_{i,j}+\bm{S}_{i,j+1}\right)\cdot }\bm{Y}_{i,j}}\prod_{i,j,\tau}\left(P_{i,j}^{\tau}+e^{-B}Q_{i,j}^{\tau}\right)\Biggr]  \notag \\
       =&\mathsf{Tr}\Biggr[\prod_{i,j}e^{-K {\left(\bm{S}_{i,j}+\bm{S}_{i+1,j}\right)\cdot\bm{X}_{i,j}}+{\left(\bm{S}_{i,j}+\bm{S}_{i,j+1}\right)\cdot}\bm{Y}_{i,j}}\times \biggr\{PPP\cdots P+e^{-B}\bigl(QPP\cdots P \notag \\
       +&PQPP\cdots P+\cdots\bigl)+e^{-2B}\bigl(QQPP\cdots P+QPQP\cdots P+QPPQP\cdots P+\cdots\bigl) \notag \\
       +&\cdots+e^{-\frac{NB}{2}}PQPQ\cdots PQ+\cdots +e^{-2NB}QQQ\cdots Q\biggr\}\Biggr],
\label{eq:4}
\end{align}
where $K=J/k_{\rm{B}}T$, $B=\lambda J/k_{\rm{B}}T$, $\bm{X}_{i,j}=\bm{R}_{i,j}^{x}+\bm{T}_{i,j}^{x}$, and $\bm{Y}_{i,j}=\bm{R}_{i,j}^{y}+\bm{T}_{i,j}^{y}$. In the last line of Eq.~(\ref{eq:4}), $N$ represents the total number of lattice points.
The operators 
\begin{align}
P_{i,j}^{\tau}\equiv&\frac{1}{4}-\bm{R}_{i,j}^{\tau}\cdot\bm{T}_{i,j}^{\tau},
\label{eq:4-2} \\
Q_{i,j}^{\tau}\equiv&\frac{3}{4}+\bm{R}_{i,j}^{\tau}\cdot\bm{T}_{i,j}^{\tau}
\label{eq:4-3}
\end{align}
project the dimer $(i,j;\tau)$ onto the singlet and triplet sectors, respectively.
It should be noted that the idendidies
\begin{align}
 \left(\bm{S}_{i,j}+\bm{S}_{i+1,j}\right)\cdot \bm{X}_{i,j}P_{i,j}^x=0 \;\;\mbox{and} \;\;\left(\bm{S}_{i,j}+\bm{S}_{i,j+1}\right)\cdot \bm{Y}_{i,j}P_{i,j}^y=0
 \label{eq:4-4}
\end{align}
indicate that we can regard the four interactions $J$ as zero when the dimer $(i,j;x)$ or $(i,j;y)$ is in the singlet state.

\section{Calculation of Ground-State Energies}\label{sec:3}

In this section, we calculate the ground-state energies of the DM sate, MDTD state, ferrimagnetic state, chain ferrimagnetic state, and square ferrimagnetic state, and determine the phase boundaries. 

\subsection{DM state case}

We consider the DM subspace, which has the singlet pairs on all of the dotted lines (see Fig.~{\ref{fig:2}}(b)). The partition function in the DM subspace, $Z(DM)$, is given by 
\begin{align}
Z(DM) =&\mathsf{Tr}\Biggr[\prod_{i,j}e^{-K {\left(\bm{S}_{i,j}+\bm{S}_{i+1,j}\right)\cdot \bm{X}_{i,j}}+{\left(\bm{S}_{i,j}+\bm{S}_{i,j+1}\right)\cdot }\bm{Y}_{i,j}}\prod_{i,j,\tau}P_{i,j}^{\tau}\Biggr].
\label{eq:5}
\end{align}
Using Eq.~(\ref{eq:4-4}), we immediately find that the energy of this state is zero: $E(DM)=0$.

\subsection{MDTD state case}

Before we discuss the MDTD state, we consider the space in which only the dimer $(i,j;x)$ is in the triplet state.
The partition function of this space, $Z(Q_{i,j}^{x})$, is written by
\begin{align}
Z(Q_{i,j}^{x})=&{\rm{Tr}} \; e^{-K\left(\bm{S}_{i,j}+\bm{S}_{i+1,j}\right)\cdot \bm{X}_{i,j}-B}Q_{i,j}^x \notag \\
			=&{\rm{Tr}} \; e^{-K\left(\bm{S}_{i,j}+\bm{S}_{i+1,j}\right)\cdot \tilde{\bm{X}}_{i,j}-B},
\label{eq:6}
\end{align}
where $\tilde{\bm{X}}_{i,j}$ is a spin-1 operator.
From Eq. (\ref{eq:6}), the ground-state energy and wave function for the present sector with one triplet dimer are obtained by 
\begin{align}
E(Q_{i,j}^{x})=J(\lambda-2)
\label{eq:7}
\end{align}
and
\begin{align}
|\phi^g\rangle_{i,j}^x=\frac{1}{\sqrt{3}}
   \left(|\!\uparrow\uparrow\rangle|t^-\rangle+|\!\downarrow\downarrow\rangle|t^+\rangle
   -\frac{|\!\uparrow\downarrow\rangle+|\!\downarrow\uparrow\rangle}{\sqrt{2}}|t^0\rangle
   \right),
\label{eq:8-1}
\end{align}
where $\{|t^+\rangle, |t^0\rangle, |t^-\rangle\}$ represent the triplet states of the dimer $(i,j;x)$ and 
$\{|\!\uparrow\uparrow\rangle,\;|\!\downarrow\downarrow\rangle,\;|\!\uparrow\downarrow\rangle,\;|\!\downarrow\uparrow\rangle)\}$ 
are the Ising basis for the two edge spins in the diamond unit.
We note that the ground state in Eq. (\ref{eq:8-1}) is non-magnetic.

\begin{figure}[h]
\begin{center}
\includegraphics[width=.90\linewidth]{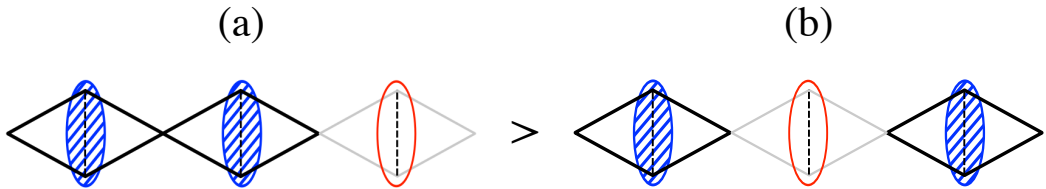}
\caption{(Color online)  Configurations where the diamonds with triplet pairs are placed continuously (a) and are placed on either side of a diamond with a singlet pair (b). Energy in the case of configuration (b) is lower than that of (a) on the basis of the variational principle.
 }
\label{fig:7}
\end{center}
\end{figure}

Equation (\ref{eq:7}) indicates that
$E(Q_{i,j}^{x})<E(DM)$ for $\lambda<2$; thus, one may think that the energy of the system decreases as the number of triplet pairs increases. 
However, the energy for a state in which two diamonds with triplet pairs are placed on either side of a diamond with a singlet pair is lower than the energy of a state in which two diamonds with triplet pairs are placed continuously, as shown in Fig.~{\ref{fig:7}},
which indicates the existance of nearest-neighbor repulsion between two diamonds with triplet pairs. This can be explained using the variational principle as shown below.

\begin{figure}[h]
\begin{center}
\includegraphics[width=.70\linewidth]{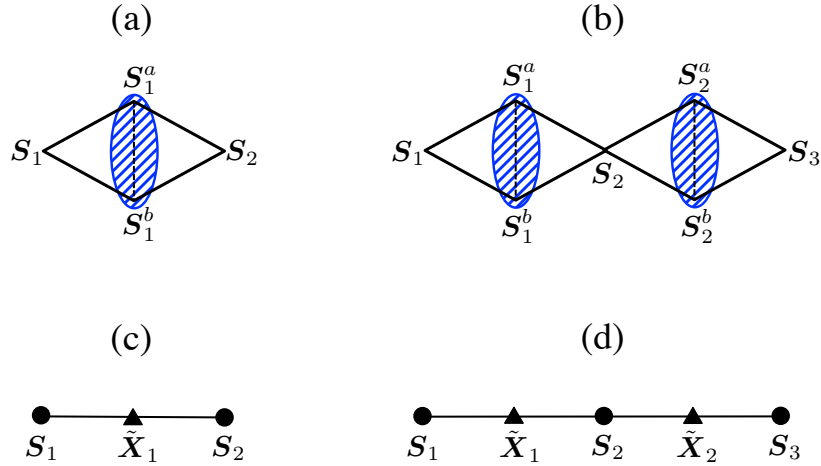}
\caption{(Color online)  (a) One diamond with a triplet pair between  $\mbox{\boldmath $S$}_1^a$ and $\mbox{\boldmath $S$}_1^b$. (b) Two diamonds with triplet pairs between $\mbox{\boldmath $S$}_1^a$, $\mbox{\boldmath $S$}_1^b$ and $\mbox{\boldmath $S$}_2^a$, $\mbox{\boldmath $S$}_2^b$ respectively. Figs. (c) and (d) are equivalent lattices to the lattice in (a) and (b), where $\tilde{\bm{X}}_{\tau}=\mbox{\boldmath $S$}_{\tau}^a+\mbox{\boldmath $S$}_{\tau}^b$ ($\tau=1,2$) is a spin-1 operator.
 }
\label{fig:8}
\end{center}
\end{figure}

We consider one diamond, and assume that $\lambda<2$ and that $\mbox{\boldmath $S$}_1^a$ and $\mbox{\boldmath $S$}_1^b$ form a triplet pair, as shown in Fig.~{\ref{fig:8}}(a). Below, we use the notation $\tilde{\bm{X}}_{\tau}=\mbox{\boldmath $S$}_{\tau}^a+\mbox{\boldmath $S$}_{\tau}^b$ where $\tau=1,2$ for a spin-1 operator. We can write 
\begin{align}
\mathcal{H}(\bm{S}_1, \tilde{\bm{X}}_1, \bm{S}_2)|\phi_g\rangle=\epsilon|\phi_g\rangle,
\label{eq:8-2}
\end{align}
where $\mathcal{H}(\bm{S}_1, \tilde{\bm{X}}_1, \bm{S}_2)$, $|\phi_g\rangle$, and $\epsilon=J(\lambda-2)$ are the Hamiltonian of one diamond, ground-state wave function, and ground-state energy respectively. Next, we consider two diamonds and assume that $\bm{S}_1^a$, $\bm{S}_1^b$ and $\bm{S}_2^a$, $\bm{S}_2^b$  form triplet pairs as shown in Fig.~{\ref{fig:8}}(b). By denoting the Hamiltonian, ground-state wave function, and ground-state energy as $\mathcal{H}'$, $|\phi'_g\rangle$, and $\epsilon'$, we obtain
\begin{align}
\epsilon'&=\langle\phi'_g|\mathcal{H}'|\phi'_g\rangle \nonumber \\
                      &=\langle\phi'_g|\mathcal{H}(\bm{S}_1, \tilde{\bm{X}}_1, \bm{S}_2)|\phi'_g\rangle+\langle\phi'_g|\mathcal{H}(\bm{S}_2, \tilde{\bm{X}}_2, \bm{S}_3)|\phi'_g\rangle \nonumber \\
                      &\ge2\epsilon,
\label{eq:9-1}
\end{align}
because $\langle\phi_g'|\mathcal{H}(\bm{S}_i, \tilde{\bm{X}}_i, \bm{S}_{i+1})|\phi_g'\rangle\ge\langle\phi_g|\mathcal{H}(\bm{S}_1, \tilde{\bm{X}}_1, \bm{S}_2)|\phi_g\rangle=\epsilon$ for $i=1, 2$ as per the variational principle.  We can show $\epsilon'\neq2\epsilon$ by using the Lieb-Mattis theorem and obtain $\epsilon'>2\epsilon$\cite{ref1,ref2}. 
Takano \textit{et al}. have presented the above discussion for a one-dimensional system, but $\epsilon'>2\epsilon$ holds for all dimensions, i.e., the nearest-neighbor repulsion between two diamonds with triplet pairs is generated for all dimensions.
Indeed, we calculate the ground-state energy of two diamonds of Fig.~{\ref{fig:8}} (b) and obtain $\epsilon'= 2J(\lambda-1.691)$. This result indicates that $\epsilon'>2\epsilon$ holds.

The relation $\epsilon'>2\epsilon$ shows that the energy in the case where two diamonds with triplet pairs are placed continuously as shown in Fig.~{\ref{fig:7}}(a) is higher than that in the case where two diamonds with triplet pairs exist independently, i.e, there is a diamond with a singlet pair between diamonds with triplet pairs as shown in  Fig.~{\ref{fig:7}}(b). Since forming a diamond with a singlet pair makes the four interactions on the solid line $J$ effectively vanish,  we can regard that individual diamonds with triplet pairs exist independently.

From the above discussion, in the case of $\lambda<2$, when the number of diamonds with triplet pairs is the largest and there is a diamond with a singlet pair between diamonds with triplet pairs, i.e., there are diamonds with singlet pairs on both sides of a diamond with a triplet pair, the system is in the ground state.
In the case of a one-dimensional system, the state in which diamonds with triplet pairs and with singlet pairs are arranged alternately becomes the ground state, and this state is the tetramer-dimer (TD) state shown in Fig.~{\ref{fig:0}} (a). 
In the case of a two-dimensional system (diamond-like decorated square lattices), we can obtain the state corresponding to the TD state. 
However, the feature that differs between the one-dimensional and two-dimensional cases is that, in the one-dimensional case, the number of configurations is only two. On the other hand, in the two-dimensional case, there are various types of configurations, and the system has macroscopic degeneracy as shown in Fig.~{\ref{fig:2}}(a). This state is the MDTD state, the energy of which is expressed as
\begin{align}
E(MDTD)=\frac{NJ}{2}(\lambda-2).
\label{eq:9-3}
\end{align}

If we regard a tetramer as dimer in the QDM, the MDTD subspace is identical to the square-lattice dimer covering states\cite{ref8,ref9}. 
It has been known that the number of dimer configurations, $N_g$, is expressed by 
\begin{align}
\lim_{N\to\infty}\frac{1}{N}N_g=0.2915609,
\label{eq:9-4}
\end{align}
for an $N=m\times n$ square lattice with torus and open boundary conditions. Eq. (\ref{eq:9-4}) corresponds to the residual entropy of the MDTD state per unit\cite{ref11}. Morita {\it et al.} have shown that the residual entropy depends on the shape of the lattice even in the bulk limit\cite{ref8}.

Thus, in the region $\lambda_{\rm c}<\lambda<2$, the MDTD state is stabilized, but in the region $\lambda<\lambda_{\rm c}$, the system can be in a ferrimagnetic state. In the case of a one-dimensional system, the system becomes a ferrimagnetic state at $\lambda_{\rm c}=0.909$\cite{ref1}. Next, we determine the phase boundary $\lambda_{\rm c}$ in the case of a two-dimensional system.

\subsection{Ferrimagnetic state case}

\begin{figure}[h]
\begin{center}
\includegraphics[width=.45\linewidth]{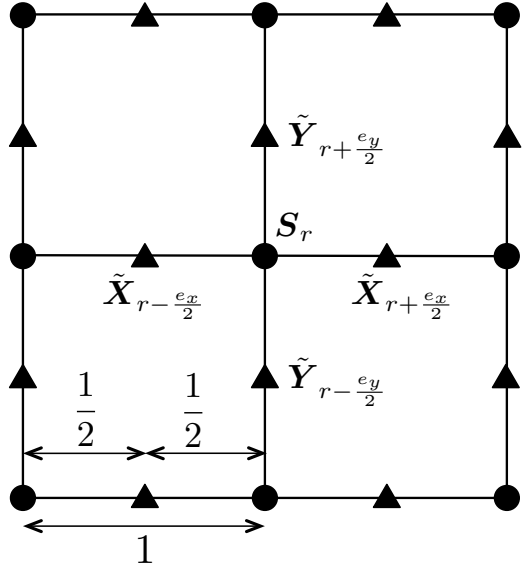}
\caption{Arrangement of spins around the $\mib{r}$ site. 
The closed circle (triangle) represents a spin $S=\frac{1}{2}$ ($\tilde{S}=1$). 
We choose the lattice spacing of this lattice as $1$.

 }
\label{fig:9}
\end{center}
\end{figure}

Using the modified spin wave method, we discuss the ferrimagnetic ground state shown in Fig.~{\ref{fig:2}}(c)
and determine the phase boundary $\lambda_{\rm c}$ between the MDTD and ferrimagnetic states. 
We redefine the spin sites as shown in Fig.~\ref{fig:9}, where $\mib{r}=(i,j)$ specifies a site for a spin $S=\frac{1}{2}$ and $\mib{r}+\rho\mib{e}_x=(i+\rho, j)$ [$\mib{r}+\rho\mib{e}_y=(i,j+\rho)$] with $\rho=\pm\frac{1}{2}$ specifies a site for a spin $S=1$ next to the site $\mib{r}$ along the $x$ [$y$] direction. 
Then, the Hamiltonian is written as 
\begin{align}
\mathcal{H}(ferri)=& J\sum_{\mib{r}}\sum_{\rho=\pm\frac{1}{2}} \left(\mbox{\boldmath $S$}_{\mib{r}}\cdot\tilde{\bm{X}}_{\mib{r}+\rho\mib{e}_x}+\mbox{\boldmath $S$}_{\mib{r}}\cdot\tilde{\bm{Y}}_{\mib{r}+\rho\mib{e}_y}\right) \notag \\
			 =& J\sum_{\mib{r}}\sum_{\rho=\pm 1/2}\Biggr[S_{\mib{r}}^z\tilde{X}_{\mib{r}+\rho\mib{e}_x}^z+\frac{1}{2}\left(S_{\mib{r}}^+\tilde{X}_{\mib{r}+\rho\mib{e}_x}^-+S_{\mib{r}}^-\tilde{X}_{\mib{r}+\rho\mib{e}_x}^+\right) \notag \\
&\qquad\qquad+S_{\mib{r}}^z\tilde{Y}_{\mib{r}+\rho\mib{e}_y}^z+\frac{1}{2}\left(S_{\mib{r}}^+\tilde{Y}_{\mib{r}+\rho\mib{e}_y}^-+S_{\mib{r}}^-\tilde{Y}_{\mib{r}+\rho\mib{e}_y}^+\right)\Biggr],
\label{eq:10}
\end{align}
where $\bm{S}_{\mib{r}}$ is a spin-$\frac{1}{2}$ operator and $\tilde{\bm{X}}_{\mib{r}}$, $\tilde{\bm{Y}}_{\mib{r}}$ are spin-1 operators. 

For Eq. (\ref{eq:10}), we introduce the Dyson-Maleev transformation,
\begin{align}
&S_{\mib{r}}^z=S-a_{\mib{r}}^{\dagger}a_{\mib{r}},        &&   S_{\mib{r}}^+=\sqrt{2S}\left(1-\frac{a_{\mib{r}}^{\dagger}a_{\mib{r}}}{2S}\right)a_{\mib{r}},                                                            &&S_{\mib{r}}^-=\sqrt{2S}a_{\mib{r}}^{\dagger}, \notag \\
&\tilde{X}_{\mib{r}+\rho\mib{e}_x}^z=b_{\mib{r}+\rho\mib{e}_x}^{\dagger}b_{\mib{r}+\rho\mib{e}_x}-\tilde{S},    &&   \tilde{X}_{\mib{r}+\rho\mib{e}_x}^+=\sqrt{2\tilde{S}}b_{\mib{r}+\rho\mib{e}_x}^{\dagger}\left(1-\frac{b_{\mib{r}+\rho\mib{e}_x}^{\dagger}b_{\mib{r}+\rho\mib{e}_x}}{2\tilde{S}}\right),  &&\tilde{X}_{\mib{r}+\rho\mib{e}_x}^-=\sqrt{2\tilde{S}} b_{\mib{r}+\rho\mib{e}_x}, \notag \\
&\tilde{Y}_{\mib{r}+\rho\mib{e}_y}^z=b_{\mib{r}+\rho\mib{e}_y}^{\dagger}b_{\mib{r}+\rho\mib{e}_y}-\tilde{S},    &&   \tilde{Y}_{\mib{r}+\rho\mib{e}_y}^+=\sqrt{2\tilde{S}}b_{\mib{r}+\rho\mib{e}_y}^{\dagger}\left(1-\frac{b_{\mib{r}+\rho\mib{e}_y}^{\dagger}b_{\mib{r}+\rho\mib{e}_y}}{2\tilde{S}}\right),   &&\tilde{Y}_{\mib{r}+\rho\mib{e}_y}^-=\sqrt{2\tilde{S}} b_{\mib{r}+\rho\mib{e}_y}, 
\label{eq:12}
\end{align}  
where $a_{\mib{r}}$ and $b_{\mib{r}+\rho\mib{e}_z}$ ($z=x$ or $y$) are Bose operators satisfying
\begin{subequations}
	\begin{eqnarray}
&\Bigr[a_{\mib{r}}, a_{\mib{r'}}^{\dagger}\Bigr]=\delta_{\mib{r},\mib{r}'},  \\
&\Bigr[b_{\mib{r}+\rho\mib{e}_z}, b_{\mib{r}'+\rho'\mib{e}_z}^{\dagger}\Bigr]=\delta_{\mib{r},\mib{r}'}\delta_{\rho,\rho'}, 
 	\end{eqnarray}
	 \label{eq:13}
\end{subequations}
and zero otherwise.

By substituting Eq. (\ref{eq:12}) into Eq. (\ref{eq:10}), we obtain 
\begin{align}
\mathcal{H}(ferri)&= J\sum_{\mib{r}}\sum_{\rho=\pm 1/2}\Biggr[-2S\tilde{S}+2\tilde{S}a_{\mib{r}}^{\dagger}a_{\mib{r}}+S\left(b_{\mib{r}+\rho\mib{e}_x}^{\dagger}b_{\mib{r}+\rho\mib{e}_x}+b_{\mib{r}+\rho\mib{e}_y}^{\dagger}b_{\mib{r}+\rho\mib{e}_y}\right) \notag \\
+&\sqrt{S\tilde{S}}\left(a_{\mib{r}}b_{\mib{r}+\rho\mib{e}_x}+a_{\mib{r}}^{\dagger}b_{\mib{r}+\rho\mib{e}_x}^{\dagger}+a_{\mib{r}}b_{\mib{r}+\rho\mib{e}_y}+a_{\mib{r}}^{\dagger}b_{\mib{r}+\rho\mib{e}_y}^{\dagger}\right)-a_{\mib{r}}^{\dagger}a_{\mib{r}}b_{\mib{r}+\rho\mib{e}_x}^{\dagger}b_{\mib{r}+\rho\mib{e}_x}-a_{\mib{r}}^{\dagger}a_{\mib{r}}b_{\mib{r}+\rho\mib{e}_y}^{\dagger}b_{\mib{r}+\rho\mib{e}_y} \notag \\
-&\frac{1}{2}\sqrt{\frac{\tilde{S}}{S}}\left(a_{\mib{r}}^{\dagger}a_{\mib{r}}a_{\mib{r}}b_{\mib{r}+\rho\mib{e}_x}+a_{\mib{r}}^{\dagger}b_{\mib{r}+\rho\mib{e}_x}^{\dagger}b_{\mib{r}+\rho\mib{e}_x}^{\dagger}b_{\mib{r}+\rho\mib{e}_x}+a_{\mib{r}}^{\dagger}a_{\mib{r}}a_{\mib{r}}b_{\mib{r}+\rho\mib{e}_y}+a_{\mib{r}}^{\dagger}b_{\mib{r}+\rho\mib{e}_y}^{\dagger}b_{\mib{r}+\rho\mib{e}_y}^{\dagger}b_{\mib{r}+\rho\mib{e}_y}\right)\Biggr].
\label{eq:14}
\end{align}

We use the mean-field approximation for the forth order terms in Eq. (\ref{eq:14}) as follows:
\begin{align}
a_{1}^{\dagger}a_{1}b_{2}^{\dagger}b_{2}\Rightarrow \langle a_{1}^{\dagger}a_{1}\rangle b_{2}^{\dagger}b_{2}
                                                        +a_{1}^{\dagger}a_{1}\langle b_{2}^{\dagger}b_{2}\rangle&
                                                        -\langle a_{1}^{\dagger}a_{1}\rangle \langle b_{2}^{\dagger}b_{2}\rangle 
                                                        +\langle a_{1}^{\dagger}b_{2}^{\dagger}\rangle a_{1}b_{2}
                                                      +a_{1}^{\dagger}b_{2}^{\dagger}\langle a_{1}b_{2}\rangle
                                                        -\langle a_{1}^{\dagger}b_{2}^{\dagger}\rangle \langle a_{1}b_{2}\rangle,  
                                                       \label{eq:15-1} 
\end{align}
\begin{align}                                                       
&a_{1}^{\dagger}a_{1}a_{1}b_{2}\Rightarrow 2\left(\langle a_{1}^{\dagger}a_{1}\rangle a_{1}b_{2}
                                                                                              +a_{1}^{\dagger}a_{1}\langle a_{1}b_{2}\rangle
                                                                                              -\langle a_{1}^{\dagger}a_{1}\rangle\langle a_{1}b_{2}\rangle\right), 
                                                                                              \label{eq:15-2}
\end{align}
\begin{align}                                                                                              
&a_{1}^{\dagger}b_{2}^{\dagger}b_{2}^{\dagger}b_{2}\Rightarrow2\left(\langle a_{1}^{\dagger}b_{2}^{\dagger}\rangle b_{2}^{\dagger}b_{2}
                                                                                                          +a_{1}^{\dagger}b_{2}^{\dagger}\langle b_{2}^{\dagger}b_{2}\rangle
                                                                                                          -\langle a_{1}^{\dagger}b_{2}^{\dagger}\rangle\langle b_{2}^{\dagger}b_{2}\rangle\right),
	 \label{eq:15-3}
\end{align}
where we have assumed $\langle a_{1}^{\dagger}b_{2}\rangle=\langle a_{1}b_{2}^{\dagger}\rangle=\langle a_{1}a_{1}\rangle=\langle b_{2}^{\dagger}b_{2}^{\dagger}\rangle=0$. 
The indexes 1 and 2 correspond to the sites $\mib{r}$ and $\mib{r}+\rho\mib{e}_x$ or $\mib{r}+\rho\mib{e}_y$, respectively. 
Furthermore, we assume that 
\begin{subequations}
	\begin{eqnarray}
&\langle a_{\mib{r}}^{\dagger}a_{\mib{r}}\rangle=S,  \\
&\langle b_{\mib{r}+\rho\mib{e}_x}^{\dagger}b_{\mib{r}+\rho\mib{e}_x}\rangle=\langle b_{\mib{r}+\rho\mib{e}_y}^{\dagger}b_{\mib{r}+\rho\mib{e}_y}\rangle=\tilde{S}
	\end{eqnarray}
	\label{eq:18}
\end{subequations}	
are true, which are the same as the constraint of the vanishing sublattice magnetizations introduced in Takahashi's modified spin wave theory\cite{ref10}. By substituting the replacements in Eqs. (\ref{eq:15-1})-(\ref{eq:15-3}) and Eqs. (\ref{eq:18}-a), (\ref{eq:18}-b) to Eq. (\ref{eq:14}), we get the mean-field Hamiltonian for $S=\frac{1}{2}$ and $\tilde{S}=1$ in the following form:
\begin{align}
\mathcal{H}^{\bold{MF}}= J&\sum_{\mib{r}}\sum_{\rho=\pm 1/2}\Biggr[2\sqrt{2}fa_{\mib{r}}^{\dagger}a_{\mib{r}}+\frac{f}{\sqrt{2}}\left(b_{\mib{r}+\rho\mib{e}_x}^{\dagger}b_{\mib{r}+\rho\mib{e}_x}+b_{\mib{r}+\rho\mib{e}_y}^{\dagger}b_{\mib{r}+\rho\mib{e}_y}\right) \notag \\
				 &+f\left(a_{\mib{r}}b_{\mib{r}+\rho\mib{e}_x}+a_{\mib{r}}^{\dagger}b_{\mib{r}+\rho\mib{e}_x}^{\dagger}+a_{\mib{r}}b_{\mib{r}+\rho\mib{e}_y}+a_{\mib{r}}^{\dagger}b_{\mib{r}+\rho\mib{e}_y}^{\dagger}\right)+2f^2-2\sqrt{2}f\Biggr] \notag \\
				 &-\mu_a\sum_{\mib{r}}\left(\frac{1}{2}-a_{\mib{r}}^{\dagger}a_{\mib{r}}\right)-\mu_b\sum_{\mib{r}}\biggr[\left(b_{\mib{r}+\rho\mib{e}_x}^{\dagger}b_{\mib{r}+\rho\mib{e}_x}-1\right)+\left(b_{\mib{r}+\rho\mib{e}_y}^{\dagger}b_{\mib{r}+\rho\mib{e}_y}-1\right)\biggr],
\label{eq:19}				 		 
\end{align}
where $\mu_a$ and $\mu_b$ are Lagrange multipliers to guarantee that Eqs. (\ref{eq:18}-a), (\ref{eq:18}-b) hold, and $f$ denotes 
\begin{align}
f\equiv-\langle a_{\mib{r}}b_{\mib{r}+\rho\mib{e}_x}\rangle=-\langle a_{\mib{r}}^{\dagger}b_{\mib{r}+\rho\mib{e}_x}^{\dagger}\rangle-\langle a_{\mib{r}}b_{\mib{r}+\rho\mib{e}_y}\rangle=-\langle a_{\mib{r}}^{\dagger}b_{\mib{r}+\rho\mib{e}_y}^{\dagger}\rangle.
\label{eq:20}
\end{align}

\begin{figure}[h]
\begin{center}
\includegraphics[width=.70\linewidth]{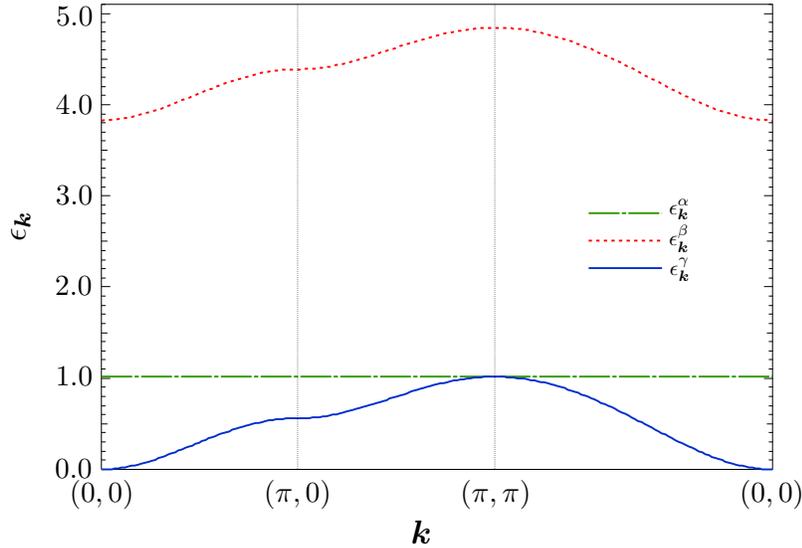}
\caption{Spin-wave dispersion for the ferrimagnetic state in the first Brillouin zone. The dashed, dotted, and solid lines represent the spectra for $\epsilon_{\bm{k}}^{\alpha}, \epsilon_{\bm{k}}^{\beta},$ and $\epsilon_{\bm{k}}^{\gamma}$ respectively.
}
\label{fig:9-2}
\end{center}
\end{figure}

We perform the Fourier transformation of $a_{\mib{r}}, b_{\mib{r}+\rho\mib{e}_x}$, and $b_{\mib{r}+\rho\mib{e}_y}$:
\begin{align}
a_{\mib{r}}=&\frac{1}{\sqrt{N}}\sum_{\mib{k}}e^{i\mib{k}\cdot\mib{r}}a_{\mib{k}}, 
\label{eq:21-1}\\
b_{\mib{r}+\rho\mib{e}_x}^{\dagger}=&\frac{1}{\sqrt{N}}\sum_{\mib{k}}e^{i\mib{k}\cdot(\mib{r}+\rho\mib{e}_x)}b_{\mib{k}}^{x\dagger}, 
\label{eq:21-2}\\
b_{\mib{r}+\rho\mib{e}_y}^{\dagger}=&\frac{1}{\sqrt{N}}\sum_{\mib{k}}e^{i\mib{k}\cdot(\mib{r}+\rho\mib{e}_y)}b_{\mib{k}}^{y\dagger}.
\label{eq:21-3}
\end{align}
Then the mean-field Hamiltonian is rewritten as follows:
\begin{align}
\mathcal{H}^{\bold{MF}}=E_0+&\sum_{\mib{k}}\epsilon_{a}a_{\mib{k}}^{\dagger}a_{\mib{k}}+\sum_{\bm{k}}\epsilon_{b}\left(b_{\mib{k}}^{x\dagger}b_{\mib{k}}^x+b_{\mib{k}}^{y\dagger}b_{\mib{k}}^y\right) \notag \\
+&2Jf\sum_{\bm{k}}\left[\cos{\frac{k_x}{2}}\left(a_{\mib{k}}b_{\bm{k}}^x+a_{\mib{k}}^{\dagger}b_{\mib{k}}^{x\dagger}\right)+\cos{\frac{k_y}{2}}\left(a_{\mib{k}}b_{\bm{k}}^y+a_{\mib{k}}^{\dagger}b_{\mib{k}}^{y\dagger}\right)\right],
\label{eq:22} 
\end{align}
where $E_0$, $\epsilon_{a}$, and $\epsilon_{b}$ are defined by
\begin{align}
	E_0=&N\left[-\frac{\mu_a}{2}+2\mu_b+4J\left(f^2-\sqrt2f\right)\right],
	\label{eq:23-1} \\
	\epsilon_{a}=&4\sqrt2Jf+\mu_a, 
	\label{eq:23-2} \\
        \epsilon_{b}=&\sqrt2Jf-\mu_b.
	\label{eq:23-3}
\end{align}
From Eq. ({\ref{eq:22}}), the Heisenberg equations for $a_{\mib{k}}, b_{\mib{k}}^{x\dagger}$, and $b_{\mib{k}}^{y\dagger}$ are written by
\begin{align}
	i\hbar\frac{\partial a_{\mib{k}}}{\partial t}=&\Bigr[a_{\mib{k}}, \mathcal{H}^{\bold{MF}}\Bigl]=\epsilon_{a}a_{\mib{k}}+2Jf\left(\cos{\frac{k_x}{2}}b_{\mib{k}}^{x\dagger}+\cos{\frac{k_y}{2}}b_{\mib{k}}^{y\dagger}\right),
	\label{eq:24-1} \\
	i\hbar\frac{\partial b_{\mib{k}}^{x\dagger}}{\partial t}=&\Bigr[b_{\mib{k}}^{x\dagger}, \mathcal{H}^{\bold{MF}}\Bigl]=-\epsilon_{b}b_{\mib{k}}^{x\dagger}-2Jf\cos{\frac{k_x}{2}}a_{\mib{k}}, 
	\label{eq:24-2}\\
	i\hbar\frac{\partial b_{\mib{k}}^{y\dagger}}{\partial t}=&\Bigr[b_{\mib{k}}^{y\dagger}, \mathcal{H}^{\bold{MF}}\Bigl]=-\epsilon_{b}b_{\mib{k}}^{y\dagger}-2Jf\cos{\frac{k_y}{2}}a_{\mib{k}}.
	\label{eq:24-3}
\end{align}
To diagonalize the mean-field Hamiltonian in Eq. ({\ref{eq:22}}), we consider the Heisenberg equations:
\begin{align}
i\hbar\frac{\partial A_{\mib{k}}}{\partial t}=\Bigr[A_{\mib{k}}, \mathcal{H}^{\bold{MF}}\Bigl]=E_{\mib{k}}A_{\mib{k}},
\label{eq:25}
\end{align}
where $A_{\mib{k}}$ is defined by
\begin{align}
A_{\mib{k}}=u_{\mib{k}}a_{\mib{k}}+v_{\mib{k}}b_{\mib{k}}^{x\dagger}+w_{\mib{k}}b_{\mib{k}}^{y\dagger},
\label{eq:26}
\end{align}
satisfying the commutation relation:
\begin{align}
\Bigr[A_{\mib{k}}, A_{\mib{k'}}^{\dagger}\Bigl]=\delta_{\mib{k}, \mib{k'}}.
\label{eq:27}
\end{align}
The eigenvalues, $E_{\mib{k}}$, and eigenvectors, $A_{\mib{k}}$, for $(u_{\mib{k}}, v_{\mib{k}}, w_{\mib{k}})\neq(0, 0, 0)$ are determined from the determinant of the following matrix:
\begin{align}
\begin{array}{|ccc|}
\epsilon_{a}-E_{\mib{k}} & -2Jf\cos{\frac{k_x}{2}}      & -2Jf\cos{\frac{k_y}{2}} \\
2Jf\cos{\frac{k_x}{2}}	     & -\epsilon_{b}-E_{\mib{k}}  & 0 \\
2Jf\cos{\frac{k_y}{2}}     &   0					& -\epsilon_{b}-E_{\mib{k}} \\
\end{array}
=0.
\label{eq:28}
\end{align}
By solving this equation, we obtain 
\begin{align}
	&\alpha_{\mib{k}}^{\dagger}\equiv\frac{\cos{\frac{k_y}{2}}}{g_{\mib{k}}}b_{\mib{k}}^{x\dagger}-\frac{\cos{\frac{k_x}{2}}}{g_{\mib{k}}}b_{\mib{k}}^{y\dagger}
	\label{eq:29-1} 
\end{align}
for $E_{\mib{k}}=-\epsilon_{b}$,
\begin{align}	
	&\beta_{\mib{k}}\equiv\sqrt{\frac{\epsilon_{a}+\epsilon_{b}+F_{\mib{k}}}{2F_{\mib{k}}}}a_{\mib{k}}+\frac{2\sqrt{2}Jf\cos{\frac{k_x}{2}}}{\sqrt{F_{\mib{k}}(\epsilon_{a}+\epsilon_{b}+F_{\mib{k}})}}b_{\mib{k}}^{x\dagger}+\frac{2\sqrt{2}Jf\cos{\frac{k_y}{2}}}{\sqrt{F_{\mib{k}}(\epsilon_{a}+\epsilon_{b}+F_{\mib{k}})}}b_{\mib{k}}^{y\dagger}
	\label{eq:29-2}
\end{align}
for $E_{\mib{k}}=\frac{\epsilon_{a}-\epsilon_{b}+F_{\mib{k}}}{2}$,
\begin{align}	
	&\gamma_{\mib{k}}^{\dagger}\equiv\sqrt{\frac{\epsilon_{a}+\epsilon_{b}-F_{\mib{k}}}{2F_{\mib{k}}}}a_{\mib{k}}+\frac{2\sqrt{2}Jf\cos{\frac{k_x}{2}}}{\sqrt{F_{\mib{k}}(\epsilon_{a}+\epsilon_{b}-F_{\mib{k}})}}b_{\mib{k}}^{x\dagger}+\frac{2\sqrt{2}Jf\cos{\frac{k_y}{2}}}{\sqrt{F_{\mib{k}}(\epsilon_{a}+\epsilon_{b}-F_{\mib{k}})}}b_{\mib{k}}^{y\dagger}
	\label{eq:29-3}
\end{align}
for $E_{\mib{k}}=\frac{\epsilon_{a}-\epsilon_{b}-F_{\mib{k}}}{2}$, with
\begin{align}
	g_{\mib{k}}=&\sqrt{\cos^2{\frac{k_x}{2}}+\cos^2{\frac{k_y}{2}}},
	 \label{eq:30-1}\\
	F_{\mib{k}}=&\sqrt{{(\epsilon_{a}+\epsilon_{b})}^2-16J^2f^2{g_{\mib{k}}}^2}.
	\label{eq:30-2}
\end{align}
From Eqs. ({\ref{eq:29-1}})-({\ref{eq:29-3}}), we can obtain
\begin{align}
	a_{\mib{k}}=&\sqrt{\frac{\epsilon_{a}+\epsilon_{b}+F_{\mib{k}}}{2F_{\mib{k}}}}\beta_{\mib{k}}-\sqrt{\frac{\epsilon_{a}+\epsilon_{b}-F_{\mib{k}}}{2F_{\mib{k}}}}\gamma_{\mib{k}}^{\dagger}, 
	\label{eq:31-1}\\
	b_{\mib{k}}^{x\dagger}=&\frac{\cos{\frac{k_y}{2}}}{g_{\mib{k}}}\alpha_{\mib{k}}^{\dagger}-\frac{2\sqrt{2}Jf\cos{\frac{k_x}{2}}}{\sqrt{F_{\mib{k}}(\epsilon_{a}+\epsilon_{b}+F_{\mib{k}})}}\beta_{\mib{k}}+\frac{2\sqrt{2}Jf\cos{\frac{k_x}{2}}}{\sqrt{F_{\mib{k}}(\epsilon_{a}+\epsilon_{b}-F_{\mib{k}})}}\gamma_{\mib{k}}^{\dagger}, 
	\label{eq:31-2}\\
	b_{\mib{k}}^{y\dagger}=&-\frac{\cos{\frac{k_x}{2}}}{g_{\mib{k}}}\alpha_{\mib{k}}^{\dagger}-\frac{2\sqrt{2}Jf\cos{\frac{k_y}{2}}}{\sqrt{F_{\mib{k}}(\epsilon_{a}+\epsilon_{b}+F_{\mib{k}})}}\beta_{\mib{k}}+\frac{2\sqrt{2}Jf\cos{\frac{k_y}{2}}}{\sqrt{F_{\mib{k}}(\epsilon_{a}+\epsilon_{b}-F_{\mib{k}})}}\gamma_{\mib{k}}^{\dagger}.
	\label{eq:31-3}
\end{align}
By substituting Eqs. ({\ref{eq:31-1}})-({\ref{eq:31-3}}) into Eq. ({\ref{eq:22}}), we obtain the mean-field Hamiltonian in the diagonalized form:
\begin{align}
\mathcal{H}^{\bold{MF}}&=E_0+\sum_{\bm{k}}\left(\epsilon_{\bm{k}}^{\alpha} \alpha_{\bm{k}}^{\dagger}\alpha_{\bm{k}}+\epsilon_{\bm{k}}^{\beta} \beta_{\bm{k}}^{\dagger}\beta_{\bm{k}}+\epsilon_{\bm{k}}^{\gamma} \gamma_{\bm{k}}^{\dagger}\gamma_{\bm{k}}+\frac{F_{\bm{k}}-\epsilon_{a}-\epsilon_{b}}{2}\right),
\label{eq:32} 
\end{align}
where the dispersion relations $\epsilon_{\bm{k}}^{\alpha}$, $\epsilon_{\bm{k}}^{\beta}$, and $\epsilon_{\bm{k}}^{\gamma}$ are given by
\begin{align}
\epsilon_{\bm{k}}^{\alpha}=&\epsilon_{b},
	\label{eq:33-1} \\
\epsilon_{\bm{k}}^{\beta}=&\frac{F_{\bm{k}}+\epsilon_{a}-\epsilon_{b}}{2},
	\label{eq:33-2} \\
\epsilon_{\bm{k}}^{\gamma}=&\frac{F_{\bm{k}}-\epsilon_{a}+\epsilon_{b}}{2}.
	\label{eq:33-3}
\end{align}
Here, by using the final result, we plot the spin-wave dispersions, $\epsilon_{\bm{k}}^{\alpha}$, $\epsilon_{\bm{k}}^{\beta}$, and $\epsilon_{\bm{k}}^{\gamma}$ in the first Brillouin zone in Fig.~\ref{fig:9-2}. We have $\epsilon_{\bm{k}}^{\gamma}=0$ for $(k_x, k_y)=(0,0)$, which is the gapless mode, and Bose condensation occurs in the ground state, as we will discuss later.

We derive the self-consistent equations for $T\to0$. By using Eqs. ({\ref{eq:21-1}})-({\ref{eq:21-3}}) and ({\ref{eq:31-1}})-({\ref{eq:31-3}}), Eqs. ({\ref{eq:18}}a), ({\ref{eq:18}}b), and ({\ref{eq:20}}) are written by
\begin{align}
	\frac{1}{2}=&\frac{1}{N}\sum_{\bm{r}}\langle a_{\bm{r}}^{\dagger}a_{\bm{r}}\rangle=\frac{1}{N}\sum_{\bm{k}}\langle a_{\bm{k}}^{\dagger}a_{\bm{k}}\rangle \notag \\
		       =&\frac{1}{N}\sum_{\bm{k}}\left[\frac{\epsilon_{a}+\epsilon_{b}+F_{\bm{k}}}{2F_{\bm{k}}}\langle \beta_{\bm{k}}^{\dagger}\beta_{\bm{k}}\rangle+\frac{\epsilon_{a}+\epsilon_{b}-F_{\bm{k}}}{2F_{\bm{k}}}\left(\langle \gamma_{\bm{k}}^{\dagger}\gamma_{\bm{k}}\rangle+1\right)\right],
	\label{eq:34-1} 
\end{align}
\begin{align}	
	1=&\frac{1}{2N}\sum_{\bm{r}}\left(\langle b_{\bm{r}}^{x\dagger}b_{\bm{r}}^{x}\rangle+\langle b_{\bm{r}}^{y\dagger}b_{\bm{r}}^{y}\rangle\right)=\frac{1}{2N}\sum_{\bm{k}}\left(\langle b_{\bm{k}}^{x\dagger}b_{\bm{k}}^{x}\rangle+\langle b_{\bm{k}}^{y\dagger}b_{\bm{k}}^{y}\rangle\right) \notag \\
		       =&\frac{1}{2N}\sum_{\bm{k}}\left[\langle \alpha_{\bm{k}}^{\dagger}\alpha_{\bm{k}}\rangle+\frac{\epsilon_{a}+\epsilon_{b}+F_{\bm{k}}}{2F_{\bm{k}}}\langle \gamma_{\bm{k}}^{\dagger}\gamma_{\bm{k}}\rangle+\frac{\epsilon_{a}+\epsilon_{b}-F_{\bm{k}}}{2F_{\bm{k}}}\left(\langle \beta_{\bm{k}}^{\dagger}\beta_{\bm{k}}\rangle+1\right)\right],
	\label{eq:34-2} 
\end{align}
and
\begin{align}		
	f=&\frac{1}{N}\sum_{\bm{k}}\frac{g_{\bm{k}}^2Jf}{F_{\bm{k}}}\left(\langle \beta_{\bm{k}}^{\dagger}\beta_{\bm{k}}\rangle+\langle \gamma_{\bm{k}}^{\dagger}\gamma_{\bm{k}}\rangle+1\right).
	\label{eq:34-3}     
\end{align}
From Eqs. ({\ref{eq:34-1}}) and ({\ref{eq:34-2}}), we obtain 
\begin{align}	
	\frac{3}{2}=&\frac{1}{N}\sum_{\bm{k}}\left(\langle \alpha_{\bm{k}}^{\dagger}\alpha_{\bm{k}}\rangle-\langle \beta_{\bm{k}}^{\dagger}\beta_{\bm{k}}\rangle+\langle \gamma_{\bm{k}}^{\dagger}\gamma_{\bm{k}}\rangle\right), 
	\label{eq:35-1} \\
	\frac{7}{2}=&\frac{1}{N}\sum_{\bm{k}}\left[\langle \alpha_{\bm{k}}^{\dagger}\alpha_{\bm{k}}\rangle+\frac{\epsilon_{a}+\epsilon_{b}}{F_{\bm{k}}}\left(\langle \beta_{\bm{k}}^{\dagger}\beta_{\bm{k}}\rangle+\langle \gamma_{\bm{k}}^{\dagger}\gamma_{\bm{k}}\rangle+1\right)\right].
	\label{eq:35-2}
\end{align}	
Further, assuming $\epsilon_{\bm{k}}^{\alpha}=\epsilon_{b}>0$, we have 
\begin{align}
\langle \alpha_{\bm{k}}^{\dagger}\alpha_{\bm{k}}\rangle=\frac{1}{e^{\epsilon_{b}/k_{\rm{B}}T}-1} \to 0
\label{eq:36}
\end{align}
for $T\to0$, and Eqs. ({\ref{eq:35-1}}) and ({\ref{eq:35-2}}) are given by
\begin{align}	
	\frac{3}{2}=&\frac{1}{N}\sum_{\bm{k}}\left(\langle \gamma_{\bm{k}}^{\dagger}\gamma_{\bm{k}}\rangle-\langle \beta_{\bm{k}}^{\dagger}\beta_{\bm{k}}\rangle\right), 
	\label{eq:36-1} \\
	\frac{7}{2}=&\frac{1}{N}\sum_{\bm{k}}\frac{\epsilon_{a}+\epsilon_{b}}{F_{\bm{k}}}\left(\langle \beta_{\bm{k}}^{\dagger}\beta_{\bm{k}}\rangle+\langle \gamma_{\bm{k}}^{\dagger}\gamma_{\bm{k}}\rangle+1\right).
	\label{eq:36-2}
\end{align}
If we assume $\epsilon_{\bm{k}}^{\gamma}>\epsilon_{\bm{k}}^{\beta}$ $(\epsilon_{a}<\epsilon_{b})$, we have $\langle \gamma_{\bm{k}}^{\dagger}\gamma_{\bm{k}}\rangle\to0$ (for $T\to0$), which is inconsistent with Eq. ({\ref{eq:36-1}}). Thus, $\epsilon_{\bm{k}}^{\gamma}<\epsilon_{\bm{k}}^{\beta}$ $(\epsilon_{a}>\epsilon_{b})$ must be true. Therefore, we obtain $\langle \beta_{\bm{k}}^{\dagger}\beta_{\bm{k}}\rangle \to 0$, and Eq. ({\ref{eq:36-1}}) is expressed by 
\begin{align}
\frac{3}{2}=&\frac{1}{N}\sum_{\bm{k}}\langle \gamma_{\bm{k}}^{\dagger}\gamma_{\bm{k}}\rangle.
\label{eq:37}
\end{align}
Since $\epsilon_{\bm{0}}^{\gamma}<\epsilon_{\bm{k}}^{\gamma}$  for $\bm{k}\neq0$ from Eq. ({\ref{eq:33-3}}), $\langle \gamma_{\bm{k}}^{\dagger}\gamma_{\bm{k}}\rangle=0$ is true except for $\bm{k}=0$. Thus, Eq. ({\ref{eq:37}}) is rewritten as
\begin{align}
\frac{3}{2}=&\frac{1}{N}\langle \gamma_{\bm{0}}^{\dagger}\gamma_{\bm{0}}\rangle,
\label{eq:38}
\end{align}
and $\epsilon_{\bm{0}}^{\gamma}=0$, i.e., Bose condensation occurs. By substituting  Eq. ({\ref{eq:38}}) and $\langle \alpha_{\bm{k}}^{\dagger}\alpha_{\bm{k}}\rangle=\langle \beta_{\bm{k}}^{\dagger}\beta_{\bm{k}}\rangle=0$ for all $\bm{k}$ and $\langle \gamma_{\bm{k}}^{\dagger}\gamma_{\bm{k}}\rangle=0$ for $\bm{k}\neq0$ into Eq. ({\ref{eq:36-2}}), we obtain the following self-consistent equation:
\begin{align}
\frac{7}{2}=&\frac{3}{2\sqrt{1-2\eta^2}}+\frac{1}{N}{\sum_{\bm{k}}}\frac{1}{\sqrt{1-\eta^2\left(\cos^2\frac{k_x}{2}+\cos^2\frac{k_y}{2}\right)}},
\label{eq:39}
\end{align}
and Eq. ({\ref{eq:39}}) can be rewritten as 
\begin{align}
\frac{7}{2}=&\frac{1}{2\sqrt{1-2\eta^2}}+\frac{1}{4\pi^2}\int_{-\pi}^{\pi}\int_{-\pi}^{\pi}dk_{x}dk_{y}\frac{1}{\sqrt{1-\eta^2\left(\cos^2\frac{k_x}{2}+\cos^2\frac{k_y}{2}\right)}},
\label{eq:40}
\end{align}
where $\eta$ is defined by
\begin{align}
\eta=\frac{4Jf}{\epsilon_{a}+\epsilon_{b}}.
\label{eq:40-2}
\end{align}
From Eq. ({\ref{eq:40}}), we have $\eta=0.535$. Furthermore, based on the above results, Eq. ({\ref{eq:34-3}}) is solved as 
\begin{align}
f=&\frac{3\eta}{4\sqrt{1-2\eta^2}}+\frac{1}{16{\pi}^2}\int_{-\pi}^{\pi}\int_{-\pi}^{\pi}dk_{x}dk_{y}\frac{\eta\left(\cos^2\frac{k_x}{2}+\cos^2\frac{k_y}{2}\right)}{\sqrt{1-\eta^2\left(\cos^2\frac{k_x}{2}+\cos^2\frac{k_y}{2}\right)}}=0.784.
\label{eq:41}
\end{align}
From Eq. ({\ref{eq:33-3}}) and $\epsilon_{\bm{0}}^{\gamma}=0$, we obtain 
\begin{align}
\epsilon_{a}\epsilon_{b}=8J^2f^2.
\label{eq:42}
\end{align}
From Eqs. ({\ref{eq:40-2}})-({\ref{eq:42}}), we obtain $\epsilon_{a}=4.844$ and $\epsilon_{b}=1.016$. Using Eq. ({\ref{eq:42}}), the ground-state energy is obtained as
\begin{align}
\lim_{T\to0}\langle\mathcal{H}^{\bold{MF}}\rangle=-4Jf^2N=-2.461NJ.
\label{eq:43}
\end{align}
Therefore, the ground-state energy of Eq. ({\ref{eq:10}}) is eventually expressed by Eq. ({\ref{eq:43}}). Thus, the ferrimagnetic ground-state energy is obtained as
\begin{align}
\lim_{T\to0}E(ferri)=(2\lambda-2.461)NJ.
\label{eq:44}
\end{align}
From Eqs. ({\ref{eq:9-3}}) and ({\ref{eq:44}}), we obtain $\lambda_{\rm c}=0.974$.

\begin{figure}[h] 
\begin{center}
\includegraphics[width=.60\linewidth]{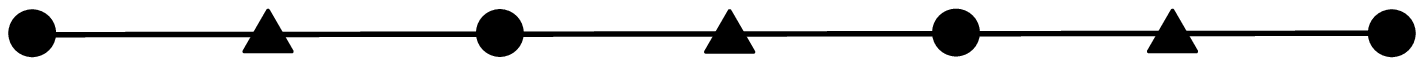}
\end{center}
\caption{Arrangement of one-dimensional system. The closed circles (triangles) represent spin-$\frac{1}{2}$ (spin-1) sites. The number of closed circles is the same as that of closed triangles. On the other hand, in the two dimensional system, the number of closed circles is half that of closed triangles (See Fig.~\ref{fig:9}).
}
\label{fig:12}
\end{figure}

In order to investigate the long-range order in the ferrimagnetic state, we consider the correlation function, $\langle\mbox{\boldmath $S_{\bm{r}}$}\cdot\mbox{\boldmath $S_{\bm{r'}}$}\rangle$, $\langle\mbox{\boldmath $S_{\bm{r}}$}\cdot\tilde{\bm{X}}_{\mib{r'}}\rangle$, $\langle\mbox{\boldmath $S_{\bm{r}}$}\cdot\tilde{\bm{Y}}_{\mib{r'}}\rangle$, $\langle\tilde{\bm{X}}_{\mib{r}}\cdot\tilde{\bm{X}}_{\mib{r'}}\rangle$, $\langle\tilde{\bm{Y}}_{\mib{r}}\cdot\tilde{\bm{Y}}_{\mib{r'}}\rangle$, and $\langle\tilde{\bm{X}}_{\mib{r}}\cdot\tilde{\bm{Y}}_{\mib{r'}}\rangle$.
For $|\bm{r}-\bm{r'}|\to \infty$, we can regard that $\langle\mbox{\boldmath $S_{\bm{r}}$}\cdot\tilde{\bm{X}}_{\mib{r'}}\rangle=\langle\mbox{\boldmath $S_{\bm{r}}$}\cdot\tilde{\bm{Y}}_{\mib{r'}}\rangle$ and $\langle\tilde{\bm{X}}_{\mib{r}}\cdot\tilde{\bm{X}}_{\mib{r'}}\rangle=\langle\tilde{\bm{X}}_{\mib{r}}\cdot\tilde{\bm{Y}}_{\mib{r'}}\rangle=\langle\tilde{\bm{Y}}_{\mib{r}}\cdot\tilde{\bm{Y}}_{\mib{r'}}\rangle$. Therefore we only have to calculate $\langle\mbox{\boldmath $S_{\bm{r}}$}\cdot\mbox{\boldmath $S_{\bm{r'}}$}\rangle$, $\langle\mbox{\boldmath $S_{\bm{r}}$}\cdot\tilde{\bm{X}}_{\mib{r'}}\rangle$, and $\langle\tilde{\bm{X}}_{\mib{r}}\cdot\tilde{\bm{X}}_{\mib{r'}}\rangle$. By using Dyson-Maleev transformation and Eq. ({\ref{eq:18}}), we obtain
\begin{align}
\langle\mbox{\boldmath $S_{\bm{r}}$}\cdot\mbox{\boldmath $S_{\bm{r'}}$}\rangle=\langle a_{\bm{r}}^{\dagger}a_{\bm{r'}}\rangle \langle a_{\bm{r'}}^{\dagger}a_{\bm{r}}\rangle.
\label{eq:45}
\end{align}
Using Eqs. ({\ref{eq:21-1}}), ({\ref{eq:31-1}}), $\langle \beta_{\bm{k}}^{\dagger}\beta_{\bm{k}}\rangle=\langle \gamma_{\bm{k}}^{\dagger}\gamma_{\bm{k}}\rangle=0$, and Eq. ({\ref{eq:38}}), Eq. ({\ref{eq:45}}) is expressed as
\begin{align}
\langle\mbox{\boldmath $S_{\bm{r}}$}\cdot\mbox{\boldmath $S_{\bm{r'}}$}\rangle={\left[\frac{3}{4}\left(\frac{1}{\sqrt{1-2\eta^2}}-1\right)+\frac{1}{8\pi^2}\int_{-\pi}^{\pi}\int_{-\pi}^{\pi}dk_{x}dk_{y}\frac{e^{i\bm{k}\cdot\bm{r''}}}{\sqrt{1-\eta^2\left(\cos^2{\frac{k_x}{2}}+\cos^2{\frac{k_y}{2}}\right)}}\right]}^2,
\label{eq:46}
\end{align}
with $\bm{r''}=\bm{r}-\bm{r'}$ If $\bm{r''}$ is sufficiently large, the integral term of the right-hand side of Eq. (\ref{eq:46}) is expressed as
\begin{align}
\int_{-\pi}^{\pi}\int_{-\pi}^{\pi}dk_{x}dk_{y}\frac{e^{i\bm{k}\cdot\bm{r''}}}{\sqrt{1-\eta^2\left(\cos^2{\frac{k_x}{2}}+\cos^2{\frac{k_y}{2}}\right)}}
	=& \int_{0}^{\infty}kdk\int_{0}^{2\pi}d{\theta}\frac{e^{ikr''\cos \theta}}{\sqrt{1-\eta^2(2-\frac{k^2}{4})}} \notag \\
	=& \frac{4\pi}{\eta}\int_{0}^{\infty}\frac{kJ_0(kr'')}{\sqrt{A^2+k^2}} \notag \\
	=&\frac{\sqrt{4}\pi e^{-Ar''}}{\eta r''},
\label{eq:47}
\end{align}
where $A=\sqrt{\frac{4(1-2\eta^2)}{2\eta^2}}$ and $J_0(kr'')$ is the Bessel function of the first kind,
\begin{align}
J_0(kr'')=\frac{1}{\pi}\int_{0}^{\pi}e^{ikr''\cos\theta}d\theta, \quad\text{and}\quad \int_{0}^{\infty}\frac{kJ_0(kr'')}{\sqrt{A^2+k^2}}dk=\frac{e^{-Ar''}}{r''}.
\label{eq:48}
\end{align}
Therefore, Eq. ({\ref{eq:47}}) becomes zero for $r''\to \infty$. By substituting $\eta=0.535$ into Eq. ({\ref{eq:46}}), we obtain 
\begin{align}
\lim_{r-r'\to\infty}\langle\mbox{\boldmath $S_{\bm{r}}$}\cdot\mbox{\boldmath $S_{\bm{r'}}$}\rangle=0.398^2=m^2.
\label{eq:49}
\end{align}
Similarly, we obtain
\begin{align}
\lim_{r-r'\to\infty}\langle\tilde{\bm{X}}_{\mib{r}}\cdot\tilde{\bm{X}}_{\mib{r'}}\rangle=0.949^2=\tilde{m}^2, 
\label{eq:50} \\
\lim_{r-r'\to\infty}\langle\mbox{\boldmath $S_{\bm{r}}$}\cdot\tilde{\bm{X}}_{\mib{r'}}\rangle=-0.378.
\label{eq:51}
\end{align}
Eqs. ({\ref{eq:49}}) and ({\ref{eq:50}}) indicate that there are magnetizations $m$ for spin-$\frac{1}{2}$ sites and $\tilde{m}$ for spin-1 sites, and Eq. ({\ref{eq:51}}) represents that spin-$\frac{1}{2}$ and spin-1 sites have opposite directions. Therefore, we conclude that the state that forms all the triplet pairs on the dotted line is a ferrimagnetic state.
Furthermore, the fact that $m=0.398<\frac{1}{2}$, $\tilde{m}=0.949<1$, and $m\tilde{m}=0.378<\frac{1}{2}$ is understood to result from the influence of spin fluctuation. 
Eqs. ({\ref{eq:49}}) and ({\ref{eq:50}}) can be also expressed, respectively, as
\begin{align}
m=\frac{1}{2}-\Delta m=0.398
\label{eq:51-2},\\
\tilde{m}=1-\Delta \tilde{m}=0.949
\label{eq:51-3}, 
\end{align}
and the ratio of spin reductions are 20 \% and 5\%, respectively. The results $\Delta m=0.102$ and  $\Delta \tilde{m}=0.051$ indicate that $\Delta m=2\Delta \tilde{m}$ holds. Thus, we can suggest that the origin of factor 2 is the ratio between the number of sites with spin-$\frac{1}{2}$ and with spin-1.
Actually, we can confirm that, in the case of a one-dimensional system (see Fig.~\ref{fig:12}), which corresponds to the ferrimagnetic ground state of the diamond chain, the number of sites with spin-$\frac{1}{2}$ is same as that with spin-1, and $\Delta m=\Delta \tilde{m}$ holds. The $n$-dimensional counterpart, $\Delta m=n\Delta \tilde{m}$, may hold, but it remains a problem for a future study. We calculate the magnetization in the ferrimagnetic ground state of the diamond chain, and obtain $m=\frac{1}{2}-\Delta m=0.291$ for spin-$\frac{1}{2}$ sites and $\tilde{m}=1-\Delta{\tilde{m}}=0.791$ for spin-1 sites  with $\Delta m=\Delta \tilde{m}=0.209$. The ratio of spin reduction of sites with spin-$\frac{1}{2}$ and with spin-1 are 42\% and 21\%, respectively. Furthermore, in the square-lattice antiferromagnet, $\Delta m=0.197$ holds and the ratio of spin reductions for spin-$\frac{1}{2}$ and spin-1 are 40\% and 20\%, respectively. This means that spin fluctuations of the square lattice antiferromagnet and of the ferrimagnetic ground state of the diamond chain are almost equal\cite{ref12}. Therefore, the results of Eqs. ({\ref{eq:51-2}}) and ({\ref{eq:51-3}}) indicate that the ground-state energy we obtained in Eq. ({\ref{eq:44}}) is more accurate than that of the square-lattice antiferromagnet and of the ferrimagnetic ground state of the diamond chain.

\subsection{Chain Ferrimagnetic State Case}

Now we consider the chain ferrimagnetic state defined in Fig.~\ref{fig:2}(d), energy of which we can obtain on the basis of results of the diamond chain.
In a diamond chain containing $M$ diamond units, the energy of the ferrimagnetic state is given by $(\lambda-1.454)MJ$.\cite{ref1}
As shown in Fig.~\ref{fig:2}(d), the diamond-like decorated square lattice with $M^2(\equiv N)$ unit cells has $M$ diamond chains; thus, the energy of the chain ferrimagnetic state is given by
\begin{align}
E(chainferri)=(\lambda-1.454)M^2J
\label{eq:52},
\end{align}
and the energy per unit cell is
\begin{align}
e(chainferri)=\frac{E(chain ferri)}{NJ}=\lambda-1.454.
\label{eq:53}
\end{align}
As we shall discuss in Sect. 4, we find that the energy in Eq. ({\ref{eq:53}}) is higher than the ground-state energy.

\subsection{Square Ferrimagnetic State Case}


Using the householder method, the energy of one-square energy is obtained as $(4\lambda-5.824)J$. Since the number of squares on a diamond-like decorated square lattice is $\frac{N}{4}$, we can obtain the energy of the square ferrimagnetic state from
\begin{align}
E(square ferri)=\frac{(4\lambda-5.824)NJ}{4}
\label{eq:54},
\end{align}
and the energy per unit cell from
\begin{align}
e(square ferri)=\frac{E(square ferri)}{NJ}=\lambda-1.456.
\label{eq:55}
\end{align}
As we shall discuss in Sect. 4, we find that Eq. ({\ref{eq:55}}) does not become the ground-state energy either.

\section{Ground State Energy and Phase Diagram}\label{sec:4}
\begin{figure}[h]
\begin{center}
\includegraphics[width=.85\linewidth]{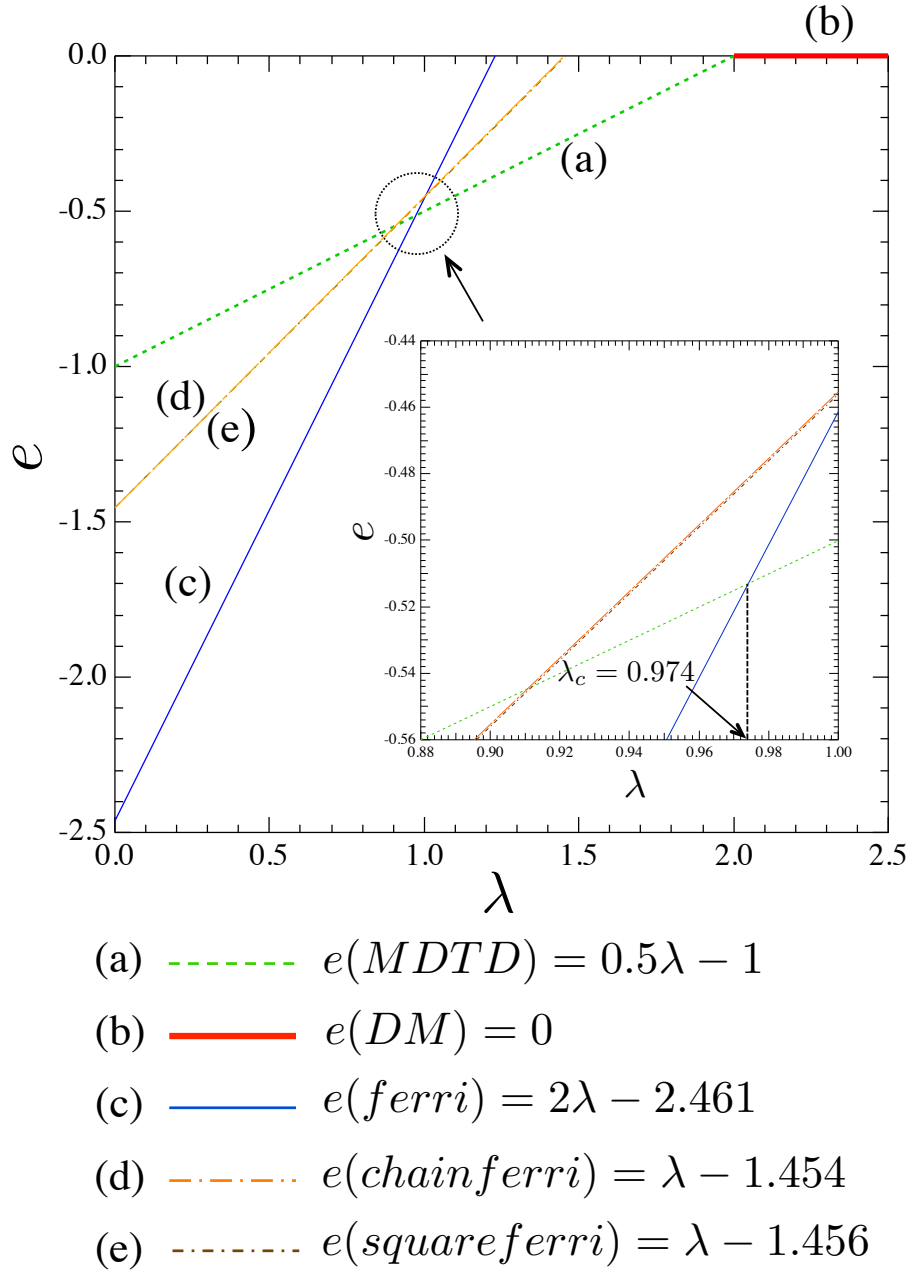}
\end{center}
\caption{(Color) Dependence of energy per unit cell $e=\frac{E}{NJ}$ on the parameter $\lambda$. $e(MDTD)$, $e(DM)$, $e(ferri)$, $e(chain ferri)$, and $e(square ferri)$ are the energies for MDTD, DM, ferrimagnetic, chain ferrimagnetic, and square ferrimagnetic states, respectively. Energy lines (a), (b), (c), (d), and (e) correspond to the states of Fig.~{\ref{fig:2}}. The inset shows an enlarged plot of intersections of each line around the point $\lambda=\lambda_{\rm c}$. }
\label{fig:10}
\end{figure}

\begin{figure}[h]
\begin{center}
\includegraphics[width=.70\linewidth]{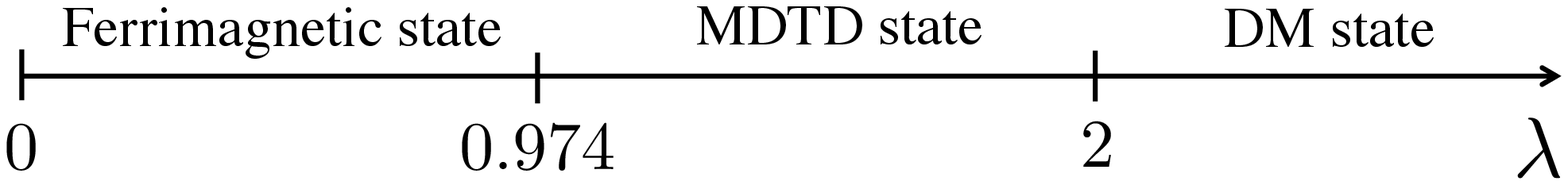}
\caption{The ground-state phase diagram of the parameter $\lambda$.    
 }
\label{fig:11}
\end{center}
\end{figure}

In Figs.~{\ref{fig:10}} and {\ref{fig:11}}, we show the dependence of the energy per unit cell $e$ on the parameter $\lambda$ and the ground-state phase diagram.
As shown in Fig. {\ref{fig:10}}, we obtain three ground-state phases: DM state for $2<\lambda$, MDTD state for $\lambda_{\rm c}<\lambda<2$, and ferrimagnetic state for $\lambda<\lambda_{\rm c}$, and we determine the phase boundary $\lambda_{\rm c}=0.974$ between the MDTD and the ferrimagnetic states. 
The obtained ground-state energies are (a) $e(MDTD)=0.5\lambda-1$, (b) $e(DM)=0$, and (c) $e(ferri)=2\lambda-2.461$ for MDTD, DM, and ferrimagnetic states, respectively. 
On the other hand, the chain ferrimagnetic and square ferrimagnetic states do not become the ground state. 
Energy lines (d) and (e), which are chain ferrimagnetic and square ferrimagnetic state energies $e(chain ferri)=\lambda-1.454$ and $e(square ferri)=\lambda-1.456$, respectively, are almost equal and a small portion of these lines is located slightly above the intersection point $\lambda=\lambda_{\rm c}$ of lines (a) and (c).


\section{Summary}\label{sec:5}

We obtained the ground-state phase diagram of the diamond-like decorated square lattice for $S=\frac{1}{2}$. 
We found three ground-state phases: DM state, MDTD state, and ferrimagnetic state, and we obtained the phase boundary $\lambda_{\rm c}=0.974$ between the MDTD and the ferrimagnetic states.
Based on the nearest-neighbor repulsion between two diamonds with triplet pairs, we have shown that the MDTD state becomes the ground state for $\lambda_{\rm c}<\lambda<2$ by using the variational principle.
By means of the modified spin wave method, we have obtained the ferrimagnetic ground-state energy and the phase boundary $\lambda_{\rm c}=0.974$. 
Furthermore, we have calculated the energy of other states, which are chain ferrimagnetic and square ferrimagnetic states, but these states do not become the ground state. 

Therefore, we can suggest that the number of ground-state phases on the diamond-like decorated square lattice is greater than three.
Judging by the result of Fig. {\ref{fig:10}}, even if new types of ground states appear in the future, we expect that they may occur near the intersection point of lines (a) and (c).

The most interesting state among the obtained ground states is the MDTD state because the MDTD subspace is identical to the Hilbert space of the Rokhsar$-$Kivelson quantum dimer model\cite{ref13}.
Very recently, we introduced further neighbor couplings and calculated the second-order effective Hamiltonian\cite{ref9}. The effective Hamiltonian is exactly the same as the square-lattice QDM with a finite hopping amplitude $t>0$ and no repulsion $v=0$, which suggests the stabilization of the plaquette phase.
However, this result is calculated under the condition that the direction of diamond units is orthogonal to the plane formed by edge spins, as in Fig. 1 of Ref. 12. On the other hand, if we choose the direction of diamond units to be parallel to the edge-spin plane, as shown in Fig. {\ref{fig:1}}, the matrix element of a perturbation bond becomes more complicated and $v/t$ may be expressed as a function of $\lambda$. It will be interesting to study the ground-state phase diagram as a function of $\lambda$. Furthermore, if a Rokhsar$-$Kivelson point appears in the function of $\lambda$, it will be very fascinating.



\end{document}